\documentclass[aps,prl,reprint,groupedaddress]{revtex4-1}

\usepackage{graphicx}
\usepackage{dcolumn}
\usepackage{bm}

\begin{document}


\title{Metal-Insulator Transition and the Role of Electron Correlation in FeO$_{2}$}


\author{Bo Gyu Jang}
\affiliation{Department of Chemistry, Pohang University of Science and Technology, Pohang 37673, Korea.}

\author{Duck Young Kim}
\email{duckyoung.kim@hpstar.ac.cn}
\affiliation{Center for High Pressure Science and Technology Advanced Research (HPSTAR), Shanghai 201203, China.}

\author{Ji Hoon Shim}
\email{jhshim@postech.ac.kr}
\affiliation{Department of Chemistry}
\affiliation{Department of Physics and Division of Advanced Nuclear Engineering, Pohang University of Science and Technology, Pohang 37673, Korea.}

\date{\today}

\begin{abstract}
Iron oxide is a key compound to understand the state of the deep Earth. It has been believed that previously known oxides such as FeO and Fe$_{2}$O$_{3}$ will be dominant at the mantle conditions. However, the recent observation of FeO$_{2}$ shed another light to the composition of the deep lower mantle (DLM) \cite{Hu2016} and thus understanding of the physical properties of FeO$_{2}$ will be critical to model DLM. Here, we report the electronic structure and structural properties of FeO$_{2}$ by using density functional theory (DFT) and dynamic mean field theory (DMFT). The crystal structure of FeO$_{2}$ is composed of Fe$^{2+}$ and O$_{2}^{2-}$ dimers, where the Fe ions are surrounded by the octahedral O atoms. We found that the bond length of O$_{2}$ dimer, which is very sensitive to the change of Coulomb interaction \emph{U} of Fe 3\emph{d} orbitals, plays an important role in determining the electronic structures. The band structures of DFT+DMFT show that the metal-insulator transition is driven by the change of \emph{U} and pressure. We suggest that the correlation effect should be considered to correctly describe the physical properties of FeO$_{2}$ compound.
\end{abstract}


\maketitle

Iron oxides are basic and important materials of the Earth’s interior. Among them, FeO and Fe$_{2}$O$_{3}$ are two end members and the most well-known compounds. However other kinds of iron oxides with new stoichiometry, such as Fe$_{4}$O$_{5}$ \cite{Lavina2011} and Fe$_{5}$O$_{6}$ \cite{Lavina2015}, are also discovered under the high pressure and temperature. Recently FeO$_{2}$, which holds an excessive amount of oxygen, is identified with both first-principles calculation and experiment near 76 GPa \cite{Hu2016}. This new iron oxide receives a great attention because it suggests an alternative scernario for describing geochemical anomalies in the lower mantle and the Great Oxidation Event. Thus, it is important to understand the correct electronic and structural properties  of FeO$_{2}$.  


FeO$_{2}$ possesses a FeS$_{2}$-type pyrite structure. The crystal structure of Fe\emph{X}$_{2}$ (\emph{X} = O or S) can be obtained by replacing \emph{X} atom in B1 type Fe\emph{X} with \emph{X}$_{2}$ dimer. FeO and FeS show a spin-state transition accompanied with Mott-type insulator to metal transition under high pressure \cite{Ohta2012, Leonov2015, Badro1999, Kobayashi2001, Rueff1999}. However, FeS$_{2}$ is a non-magnetic compound where the six Fe \emph{d} electrons occupy the \emph{t$_{2g}$} ground states \cite{Rueff1999, Fujimori1996, Chattopadhyay1985, Miyahara1968}. NiS$_{1-x}$Se$_{x}$ also has a same crystal structure with FeO$_{2}$. It exhibits a complex phase diagram including MIT and magnetic phase transition depending on composition \emph{x}, temperature, and pressure due to partially filled \emph{e$_{g}$} orbital \cite{Kunes2010, Moon2015}. Several previous studies have reported that the \emph{p} orbitals of S$_{2}$ dimer play an important role in describing electronic structures of this compounds \cite{Kunes2010, Moon2015}. So we can expect that O$_{2}$ dimer may also be an driving factor for determining electronic and physical properties of FeO$_{2}$. 

It is well known that standard density functional theory (DFT) fails to reproduce the physical properties and the electronic structures of many TMO compounds because electron correlation effect of \emph{d} orbitals cannot be described properly. Alternatively, DFT+\emph{U} which includes the correlation effect of localized orbitals such as 3\emph{d} gives better results for structural properties, magnetic moments, and electronic structures. Dynamic Mean Field Theory (DMFT) has been believed to be a more advanced technique which deals with local electronic correlation problems exactly \cite{Kotliar2004}. DMFT can describe weakly correlated electron system because it can capture both the itinerant and localized nature of spectral function. DMFT has been combined with DFT (DFT+DMFT), and it has been widely used to describe the correlated physics of real materials in a first-principles manner. 

In this paper, we investigate structure properties and electronic structure using DFT and DFT+DMFT approaches. First, the electronic structure of experimentally reported FeO$_{2}$ is calculated from DFT+DMFT. The calculated electronic structures indicate that $\sigma$* band of O$_{2}$ dimer plays an important role in determining the physical properties of this system. We also study that the correlation effect of Fe \emph{d} orbitals should be considered to describe the crystal structure of FeO$_{2}$ properly. Last, we find that MIT can occur by varying volume or changing Coulomb interaction \emph{U}. O$_{2}$ dimer bond length is a governing parameter to determine the MIT in this system which is sensitively affected by correlation strength of Fe \emph{d} orbitals.

\begin{figure}
\includegraphics[width=\linewidth]{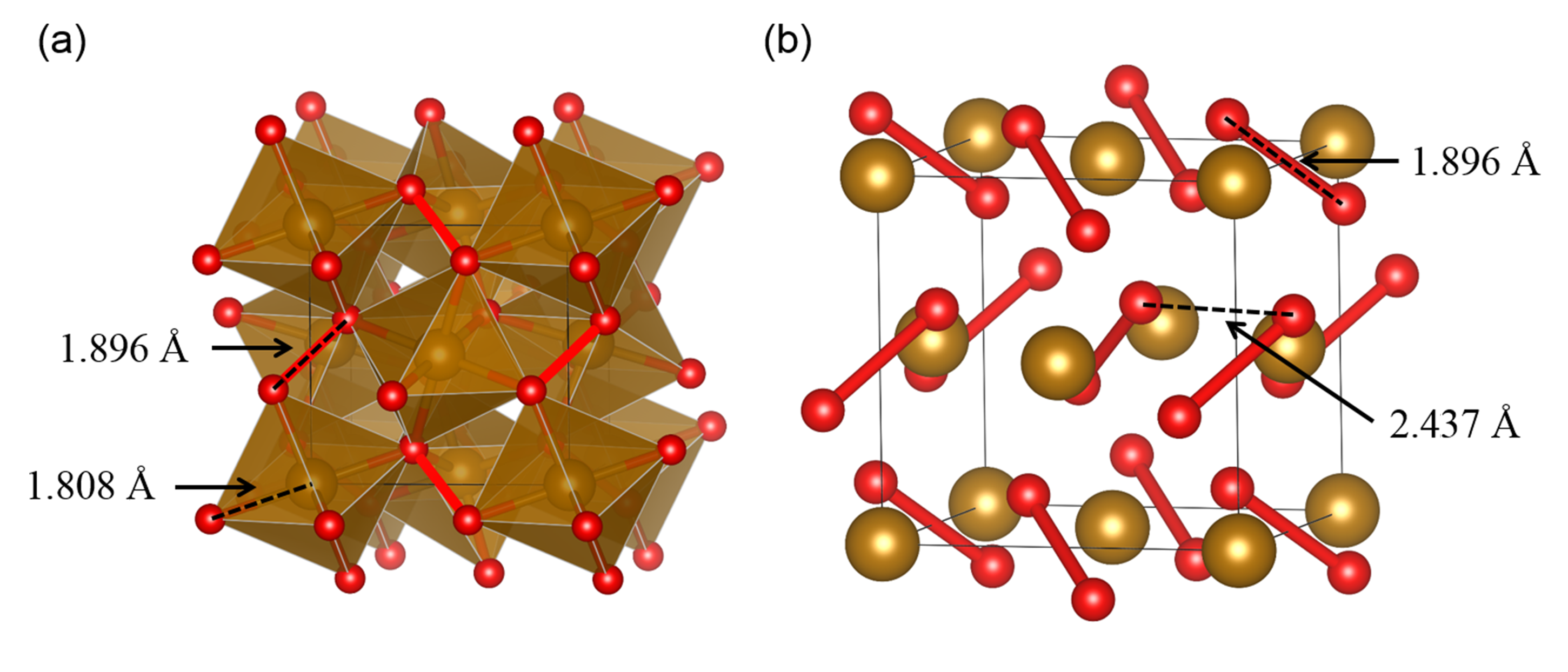}
\caption{\label{fig:epsart} Crystal structure of FeO$_{2}$. Brown and red spheres indicate Fe and O atoms, respectively. (a) Fe atom surrounded by six O atoms makes octahedral symmetry where the Fe-O bond length is 1.808 {\AA}. (b) O-O dimers in FeO$_{2}$ crystal. Fe-O bond is omitted for clarity. The bond length of O$_{2}$ dimer is 1.896 {\AA}  and the distance between second nearest O atom is 2.437 {\AA}.}
\end{figure}

DFT calculation is performed with WIEN2k code \cite{Blaha2001}, which uses a full-potential augmented plane-wave method. We use the generalized gradient approximation by Perdew, Burke, and Ernzerhof (PBE GGA) to exchange-correlation functional \cite{Perdew1997}. A 12$\times$12$\times$12 \emph{k}-points mesh is used for self-consistent calculation. Effective one electron Hamiltonian is generated from WIEN2k calculation and electronic correlation effect of Fe \emph{d} orbitals is treated by local self-energy, which is considered by using continuous time quantum Monte Carlo (CTQMC) impurity solver. The detail of DMFT implementation to the DFT method has been introduced in ref.\cite{Haule2010} explicitly. We consider a paramagnetic state at temperature T=200 K. For structural optimization at different volumes, we use the Vienna ab initio package (VASP) \cite{Kresse1996}, where a plane-wave cutoff is set to 500 eV and a 10$\times$10$\times$10 \emph{k}-points mesh is used.

Figure 1 shows the experimental crystal structure of FeO$_{2}$ at 76 GPa which contains four Fe atoms and eight O atoms in the unit cell. It possesses a simple cubic structure with a space group Pa\=3  where the four Fe atoms are located at (0, 0, 0), (0, 0.5, 0.5), (0.5, 0, 0.5), and (0.5, 0.5, 0). The eight O atoms are located at  $\pm(a, a, a),  \pm(0.5-a, -a, 0.5+a),  \pm(-a, 0.5+a, 0.5-a)$, and   $\pm(0.5+a, 0.5-a, -a)$, where $a$=0.3746 at the experimental structure. Fe atom surrounded by six O atoms makes slightly distorted octahedral symmetry where the Fe-O bond length is 1.808 {\AA}. Each octahedron shares oxygen atoms at vertex or it is connected by O-O bonding which makes O$_{2}$ dimer as shown in Fig. 1. The bond length of O$_{2}$ dimer is 1.896 {\AA} and the distance between second nearest O atoms is 2.437 {\AA}, which is quite large compared to the O$_{2}$ dimer bond length. Thus, one can expect that O$_{2}$ dimer forms  $\sigma$ and $\pi$ molecular orbitals which may play an important role in this system. 

\begin{figure} 
\centering
\includegraphics[width=\linewidth]{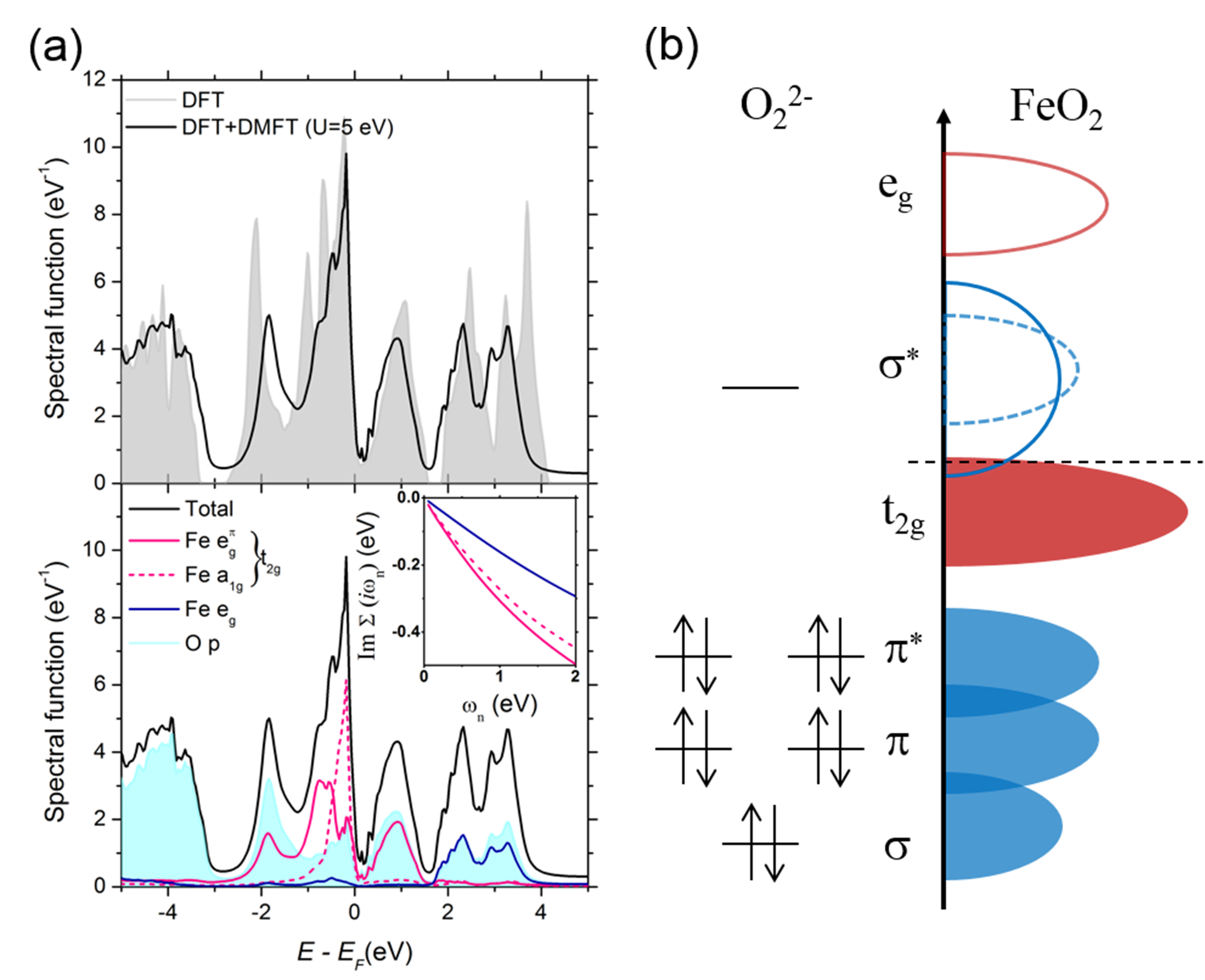}
\caption{(a) Calculated spectral functions from standard DFT and DFT+DMFT (\emph{U}=5 eV, \emph{J}=0.8 eV, and T=200 K) (upper panel) and orbital resolved spectra from DFT+DMFT (lower panel). Inset shows imaginary part of electron self-energy on Matsubara frequency for each orbitals. (b) Molecular orbital diagram of O$_{2}^{2-}$ and schematic DOS of FeO$_{2}$. FeO$_{2}$ shows metallic behavior due to the broad O$_{2}$ $\sigma$* band. 
}
\label{fig:Fig2}
\end{figure}

First, we perform calculations of the electronic structures using DFT and its combination to the DMFT (DFT+DMFT) method on the reported crystal structure of FeO$_{2}$ at 76 GPa. Figure 2 (a) displays spectral functions from standard DFT calculation and DFT+DMFT calculation. In DFT+DMFT, we use an on-site Coulomb interaction \emph{U} = 5 eV and a Hund coupling constant \emph{J} = 0.8 eV. Spectral function of DFT+DMFT calculation is slightly renormalized from that of DFT calculation, which indicates that correlation effect on spectral function is very weak. The inset figure displays the imaginary part of electron self-energy on Matsubara frequency. Fe \emph{t$_{2g}$} and \emph{e$_{g}$} bands show mass enhancement of m*/ m of $\sim$1.4 and $\sim$1.2, respectively. 

The spectral function in Fig. 2 (a) clearly shows metallic behavior of FeO$_{2}$ at 76 GPa. Its \emph{e$_{g}$} orbitals, which is well above Fermi level by $\sim$2 - 4 eV, are fully empty which lead to the low spin states resulting in the corresponding local magnetic moment to be zero. Calculated local moment in our DFT+DMFT calculation also indicates the stable non-magnetic ground state of this compound at high temperature. The \emph{t$_{2g}$} bands split into \emph{e$_{g}^{\pi}$} doublet and an \emph{a$_{1g}$} singlet due to distorted FeO$_{6}$ octahedron symmetry \cite{Kunes2010}. Overall bandwidth of \emph{t$_{2g}$} bands is around 2 eV and locates just below Fermi level while the broad $\sigma$* band of O$_{2}$ dimer is just above Fermi level with the bandwidth of around 3 eV. Thus, O$_{2}$ dimer takes two electrons from a nearest Fe atom forming hyperoxide O$_{2}^{2-}$ and electrons are occupied up to $\pi$* antibonding orbitals. A schematic electronic structures are shown in Fig. 2 (b) based on the molecular orbital energy diagram. The $\pi$ and $\sigma$ bands from O$_{2}$ dimer are completely filled making broad bands well below Fermi level. In the calculated electronic structures of FeO$_{2}$ at 76 GPa, there is an overlap between the \emph{t$_{2g}$} band and the $\sigma$* band due to their large bandwidth, and it forms a metallic ground state. We also perform DFT+DMFT calculation at a  given crystal structure with bigger \emph{U} value but band gap does not open due to the broad $\sigma$* band at Fermi level as shown in Fig. 2. Therefore, the metallic ground state is robust regardless of the size of the correlation effect. 

Note that the bandwidth and the position of $\sigma$* band is subject sensitively to the bond length of O$_{2}$ dimer. In Fig. 2 (b), a schematic DOS shows a possible insulating ground state by reducing the bandwidth of $\sigma$* bands. We speculate that this system may locate near the metal-insulator transition point, which is controlled by O$_{2}$ dimer. In the following, we investigate further the effect of pressure on the crystal structure and its effect on the electronic properties.

We investigate the change of the electronic structures with respect to volume. Before optimizing the crystal structure at several volumes, we check if standard DFT captures the experimental structure at 76 Gpa properly. However, standard DFT calculation overestimates O$_{2}$ dimer bond length by $\sim$0.2 {\AA}, which is originated from underestimated Fe-O bond length. At given volume, O$_{2}$ dimer bond length and Fe-O distance are determined by oxygen position parameter $a$ in the unit cell. As the position parameter $a$ increases, Fe-O distance increases slowly while O$_{2}$ dimer bond length decreases rapidly. When Fe-O distance increases by 0.05 {\AA}, O$_{2}$ dimer bond length decreases by $\sim$0.5 {\AA}. So O$_{2}$ dimer bond length is easily affected by small change of Fe-O bond length. DFT calculation predicts Fe-O distance to be shorter than that of experimental report and thus the corresponding O$_{2}$ dimer bond length is estimated to be longer. It is expected that the standard DFT calculation cannot describe a localized picture of Fe \emph{d} orbitals properly and overestimates the bonding strength between Fe and O. 

Furthermore, we perform DFT+\emph{U} calculation to check the change of crystal structure depending on on-site Coulomb interaction \emph{U}. We find that O$_{2}$ dimer bond length is very sensitively affected by a choice of \emph{U} value as shown in Fig. 3 (a). Although the change in Fe-O distance, which is directly affected by \emph{U}, is small, O$_{2}$ dimer bond length is rapidly changed as we discussed above. With an increase of \emph{U} value, hybridization between O$_{2}$ \emph{p} orbitals and Fe \emph{t$_{2g}$} bands decreases to make Fe-O bond length increase. O$_{2}$ dimer bond length is affected sensitively by this change and becomes closer to that of experimental value. This indicates that the correlation effect on Fe \emph{d} orbitals should be considered to describe properly the crystal structure of FeO$_{2}$, especially O$_{2}$ dimer bond length. 

O$_{2}$ dimer bond length also affects the electronic structure of FeO$_{2}$. As O$_{2}$ dimer bond length decreases, the interaction between the adjacent dimers decreases making the band width of O$_{2}$ $\sigma$* orbitals narrow. Splitting between bonding and antibonding orbitals of O$_{2}$ dimer increases by pushing up the O$_{2}$ $\sigma$* orbitals above Fermi level as shown in Fig. 2 (b). We notice that O$_{2}$ $\sigma$* bands are completely removed from Fermi level at \emph{U} = 6 eV to make a gap between Fe \emph{t$_{2g}$} bands and O$_{2}$ $\sigma$* band. FeO$_{2}$ eventually turns into an insulator as described in Fig. 2 (b) with dotted $\sigma$* band. 

\begin{figure} 
\centering
\includegraphics[width=0.8\linewidth]{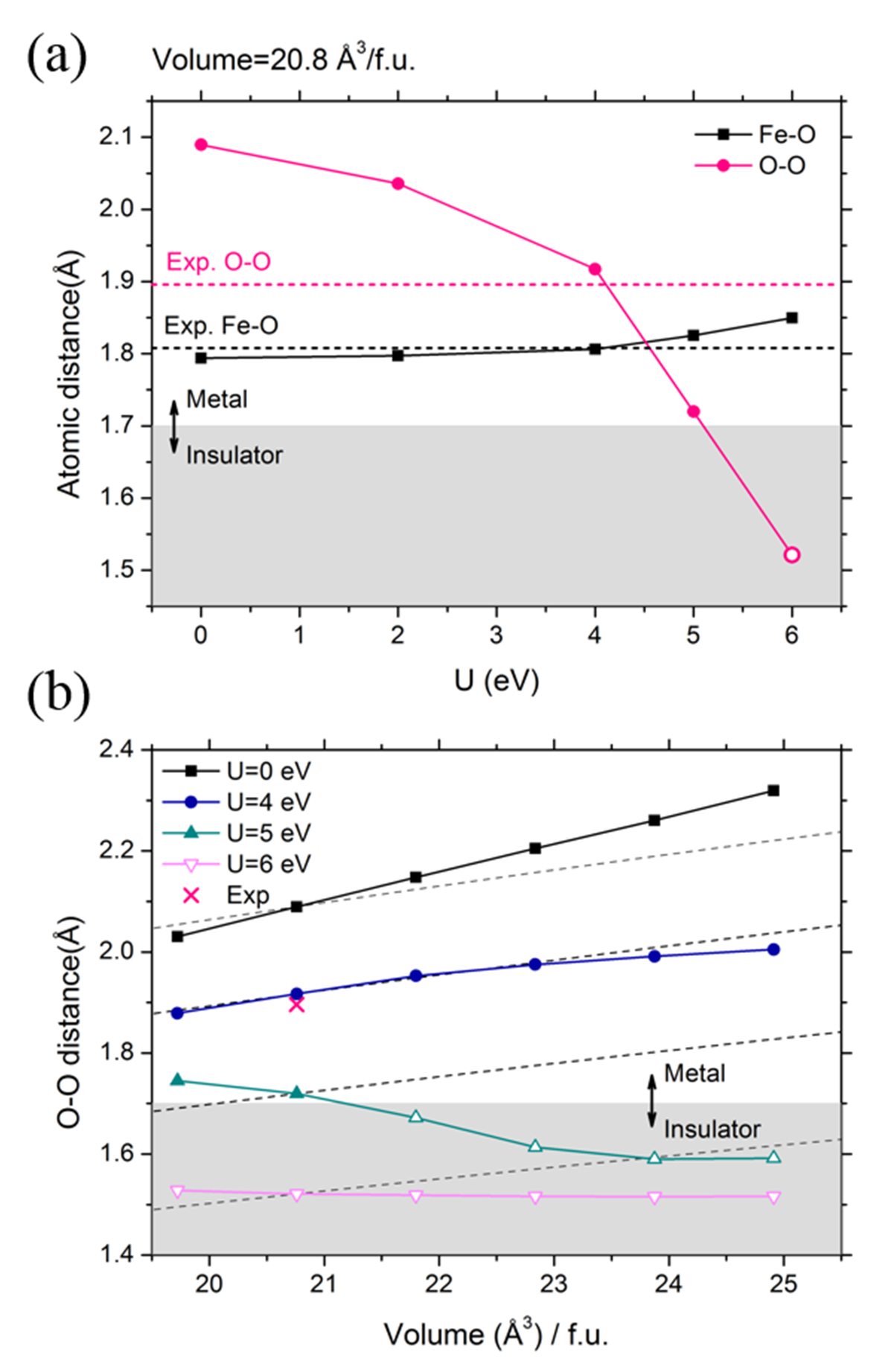}
\caption{(a) Variation of Fe-O and O-O distance in experimental volume (20.8 {\AA}$^{3}$/f.u.) at 76 Gpa with respect to Coulomb interaction energy \emph{U}. (b) Calculated O$_{2}$ dimer bond length with respect to volume for several  \emph{U} value from DFT+\emph{U} calculation. Below critical O$_{2}$ dimer bond length ($\sim$1.7 {\AA}), it turns into insulator from metal. The filled symbols and open symbols indicate the metallic and insulating states respectively.}
\label{fig:Fig3}
\end{figure}

\begin{figure*} 
\centering
\includegraphics[width=0.8\linewidth]{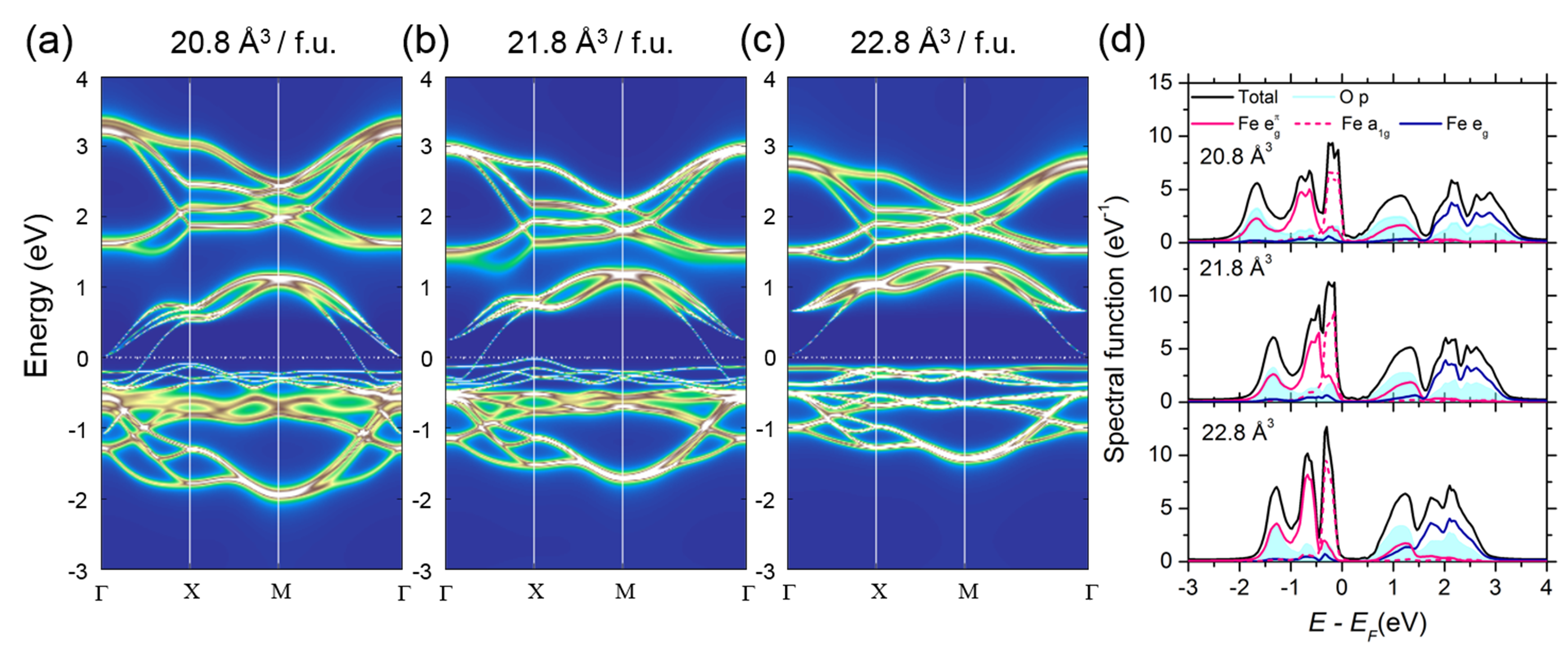}
\caption{Calculated momentum resolved spectral function for (a) 20.8 {\AA}$^{3}$ / f.u., (b) 21.8 {\AA}$^{3}$ / f.u., and (c) 22.8 {\AA}$^{3}$ / f.u. at T = 200 K. The system is metallic for (a) and (b), and insulating for (c) which makes gap between O$_{2}$  $\sigma$* band and Fe \emph{t$_{2g}$} (\emph{e$_{g}^{\pi}$} and \emph{a$_{1g}$}) band. (d) Spectral functions show metal to insulator transition clearly. Small spectral weight near Fermi level disappear as O$_{2}$  $\sigma$* band width decreases.
}
\label{fig:Fig4}
\end{figure*}

We obtain the optimized structures at several volumes with varying \emph{U} value, as shown in Fig. 3 (b). O$_{2}$ dimer bond length increases as volume increases with a choice of \emph{U} value up to 4 eV. However, this trend is reversed at higher \emph{U} value. It can be understood from a competition between Fe-O bonding strength and correlation effect of Fe 3\emph{d} orbitals. Note that stronger Fe-O bonding strength gives shorter Fe-O bonding, which results in longer O$_{2}$ dimer bond length. The gray dashed reference lines in Fig. 3 (b) exhibit a simple O$_{2}$ dimer bond length change in accordance with volume expansion without atomic position relaxation with respect to O$_{2}$ dimer bond length at 20.8 {\AA}$^{3}$. When the slope of O$_{2}$ dimer bond length with respect to volume is steeper than that of the guided line, the effect of Fe-O bonding strength predominates over the correlation effect. In this case, the O$_{2}$ dimer bonding becomes weaker with increasing volume at small \emph{U} value. Although the O$_{2}$ dimer bond length keeps increasing at \emph{U} = 4 eV, the slope is smaller than the guided line, which indicates that the formation of O$_{2}$ dimer becomes preferable at bigger volume. When \emph{U} value is larger than 5 eV, the correlation effect becomes more dominant than Fe-O bonding strength so that the formation of O$_{2}$ dimer is much preferred. We already discussed that the tiny change in Fe-O bond, which is induced by the change of \emph{U}, makes big difference in O$_{2}$ dimer bond length which can affect the electronic structure of FeO$_{2}$ significantly. Comparison of O$_{2}$ dimer bond length between experiment and theory will be a useful test to confirm the importance of the correlation effect. We suggest that exact measurements of O$_{2}$ dimer bond length with respect to volume can verify which \emph{U} value is proper for the correct description of this system. 

We find that FeO$_{2}$ turns into insulator below critical O$_{2}$ dimer bond length about 1.7 {\AA}. MIT is observed at \emph{U} = 5 eV with varying the volume as shown in Fig. 3 (b). It is also observed at constant volume by varying \emph{U} value if the change of O$_{2}$ dimer bond length is correctly captured at given \emph{U} value as shown in Fig. 3 (a). Momentum resolved spectral functions are calculated using DFT+DMFT to investigate MIT more carefully as shown in Fig. 4. In all volumes, the DFT+DMFT results always show weak correlation with mass enhancement of m*/m less than 1.5. The bandwidth of \emph{t$_{2g}$} and \emph{e$_{g}$} bands decreases from $\sim$2 eV to $\sim$1.5 eV as volume increases from 20.8 {\AA}$^{3}$ to 22.8 {\AA}$^{3}$. The crystal field splitting between \emph{t$_{2g}$} and \emph{e$_{g}$} bands also decreases from 3.5 eV to 3 eV as Fe-O bond length increases. O$_{2}$ $\sigma$* bands also show a decrease in width. The band width of $\sigma$* band is $\sim$2 eV at volume of 20.8 {\AA}$^{3}$ and decreases by $\sim$0.5 eV at volume of 22.8 {\AA}$^{3}$. The tail of $\sigma$* band gradually moves from -1 eV to above Fermi level as the band width decreases, leading to MIT. 

It is worth to note that the MIT is a band insulator type not a Mott-type because it is driven by the change of the band widths and positions of O$_{2}$ $\sigma$* and Fe \emph{t$_{2g}$} bands, which are determined by O$_{2}$ dimer bond length. As we discussed above, electronic structure is robust only with the change of \emph{U} values while O$_{2}$ dimer bond length is fixed. It is interesting that the correlation effect is very limited directly to the spectral function, but it plays an important role in the MIT through the change of the crystal structure. O$_{2}$ dimer bond length, which controls MIT in this system, is sensitively affected by the competition between correlation effect and Fe-O bond strength. 

It should be also noted that O$_{2}$ dimer bond length can be easily tuned by external condition such as chemical doping and/or oxygen vacancy which can affect the correlation strength of Fe \emph{d} orbitals \cite{Imada1998}. As we discussed above, small change of Fe-O bond length affects O$_{2}$ dimer bond length significantly. If Fe-O distance increases only by $\sim$0.02 {\AA} from the experimental distance, it easily turns into an insulator due to the sensitive change in O$_{2}$ dimer distance as shown in Fig. 3 (a). So the electronic properties can be significantly changed by external conditions. 

A stable phase of FeO$_{2}$ was observed under deep lower mantle condition with very high pressure and temperature. To simulate the mantle condition, we also investigate the electronic structure of FeO$_{2}$ at high temperature up to 2000 K. When DFT+DMFT calculations are performed on the experimental structure at 76GPa, electronic structures of FeO$_{2}$ exhibit metallic nature at any temperature of our interest. So, we expect that the O$_{2}$ dimer bond length is the most important parameter to determine the physical properties under the lower mantle condition. On the other hand, the magnetism also might be an important parameter to the electronic structures at very low temperature, although FeS$_{2}$, which has a same crystal structure, is reported to be a non-magnetic compound for whole temperature range \cite{Rueff1999, Fujimori1996, Chattopadhyay1985, Miyahara1968}. 

Using DFT and DFT+DMFT calculation, we investigated electronic structure and structural properties of FeO$_{2}$ under high pressure. Calculated spectral function from DFT+DMFT indicates that correlation effect on the electronic structure of given crystal structure is small. However, the correlation effect of Fe \emph{d} orbitals plays an important role to determine crystal structure of FeO$_{2}$. Specifically, O$_{2}$ dimer bond length is sensitively affected by the choice of \emph{U} value. We find that FeO$_{2}$ shows MIT at the critical O$_{2}$ dimer bond length of $\sim$1.7 {\AA}, which can be induced by changing the \emph{U} value or volume. We suggest that the correlation effect should be considered to describe correct structural and electronic properties of FeO$_{2}$. 

\begin{acknowledgments}
This research was supported by the National Research Foundation of Korea (NRF) grant funded by the Korea government (MSIP) (No. 2015R1A2A1A15051540), and the Supercomputing Center/Korea Institute of Science and Technology Information with supercomputing resources including technical support (KSC-2016-C1-0003). DYK acknowledges the financial support by the NSAF (U1530402).
\end{acknowledgments}

\bibliography{reference}

\begin{thebibliography}{19}%
\makeatletter
\providecommand \@ifxundefined [1]{%
 \@ifx{#1\undefined}
}%
\providecommand \@ifnum [1]{%
 \ifnum #1\expandafter \@firstoftwo
 \else \expandafter \@secondoftwo
 \fi
}%
\providecommand \@ifx [1]{%
 \ifx #1\expandafter \@firstoftwo
 \else \expandafter \@secondoftwo
 \fi
}%
\providecommand \natexlab [1]{#1}%
\providecommand \enquote  [1]{``#1''}%
\providecommand \bibnamefont  [1]{#1}%
\providecommand \bibfnamefont [1]{#1}%
\providecommand \citenamefont [1]{#1}%
\providecommand \href@noop [0]{\@secondoftwo}%
\providecommand \href [0]{\begingroup \@sanitize@url \@href}%
\providecommand \@href[1]{\@@startlink{#1}\@@href}%
\providecommand \@@href[1]{\endgroup#1\@@endlink}%
\providecommand \@sanitize@url [0]{\catcode `\\12\catcode `\$12\catcode
  `\&12\catcode `\#12\catcode `\^12\catcode `\_12\catcode `\%12\relax}%
\providecommand \@@startlink[1]{}%
\providecommand \@@endlink[0]{}%
\providecommand \url  [0]{\begingroup\@sanitize@url \@url }%
\providecommand \@url [1]{\endgroup\@href {#1}{\urlprefix }}%
\providecommand \urlprefix  [0]{URL }%
\providecommand \Eprint [0]{\href }%
\providecommand \doibase [0]{http://dx.doi.org/}%
\providecommand \selectlanguage [0]{\@gobble}%
\providecommand \bibinfo  [0]{\@secondoftwo}%
\providecommand \bibfield  [0]{\@secondoftwo}%
\providecommand \translation [1]{[#1]}%
\providecommand \BibitemOpen [0]{}%
\providecommand \bibitemStop [0]{}%
\providecommand \bibitemNoStop [0]{.\EOS\space}%
\providecommand \EOS [0]{\spacefactor3000\relax}%
\providecommand \BibitemShut  [1]{\csname bibitem#1\endcsname}%
\let\auto@bib@innerbib\@empty
\bibitem [{\citenamefont {Hu}\ \emph {et~al.}(2016)\citenamefont {Hu},
  \citenamefont {Kim}, \citenamefont {Yang}, \citenamefont {Yang},
  \citenamefont {Meng}, \citenamefont {Zhang},\ and\ \citenamefont
  {Mao}}]{Hu2016}%
  \BibitemOpen
  \bibfield  {author} {\bibinfo {author} {\bibfnamefont {Q.}~\bibnamefont
  {Hu}}, \bibinfo {author} {\bibfnamefont {D.~Y.}\ \bibnamefont {Kim}},
  \bibinfo {author} {\bibfnamefont {W.}~\bibnamefont {Yang}}, \bibinfo {author}
  {\bibfnamefont {L.}~\bibnamefont {Yang}}, \bibinfo {author} {\bibfnamefont
  {Y.}~\bibnamefont {Meng}}, \bibinfo {author} {\bibfnamefont {L.}~\bibnamefont
  {Zhang}}, \ and\ \bibinfo {author} {\bibfnamefont {H.-K.}\ \bibnamefont
  {Mao}},\ }\href {\doibase 10.1038/nature18018} {\bibfield  {journal}
  {\bibinfo  {journal} {Nature}\ }\textbf {\bibinfo {volume} {534}},\ \bibinfo
  {pages} {241} (\bibinfo {year} {2016})}\BibitemShut {NoStop}%
\bibitem [{\citenamefont {Lavina}\ \emph {et~al.}(2011)\citenamefont {Lavina},
  \citenamefont {Dera}, \citenamefont {Kim}, \citenamefont {Meng},
  \citenamefont {Downs}, \citenamefont {Weck}, \citenamefont {Sutton},\ and\
  \citenamefont {Zhao}}]{Lavina2011}%
  \BibitemOpen
  \bibfield  {author} {\bibinfo {author} {\bibfnamefont {B.}~\bibnamefont
  {Lavina}}, \bibinfo {author} {\bibfnamefont {P.}~\bibnamefont {Dera}},
  \bibinfo {author} {\bibfnamefont {E.}~\bibnamefont {Kim}}, \bibinfo {author}
  {\bibfnamefont {Y.}~\bibnamefont {Meng}}, \bibinfo {author} {\bibfnamefont
  {R.~T.}\ \bibnamefont {Downs}}, \bibinfo {author} {\bibfnamefont {P.~F.}\
  \bibnamefont {Weck}}, \bibinfo {author} {\bibfnamefont {S.~R.}\ \bibnamefont
  {Sutton}}, \ and\ \bibinfo {author} {\bibfnamefont {Y.}~\bibnamefont
  {Zhao}},\ }\href {\doibase 10.1073/pnas.1107573108} {\bibfield  {journal}
  {\bibinfo  {journal} {Proc. Natl. Acad. Sci. U. S. A}\ }\textbf {\bibinfo
  {volume} {108}},\ \bibinfo {pages} {17281} (\bibinfo {year}
  {2011})}\BibitemShut {NoStop}%
\bibitem [{\citenamefont {Lavina}\ and\ \citenamefont
  {Meng}(2015)}]{Lavina2015}%
  \BibitemOpen
  \bibfield  {author} {\bibinfo {author} {\bibfnamefont {B.}~\bibnamefont
  {Lavina}}\ and\ \bibinfo {author} {\bibfnamefont {Y.}~\bibnamefont {Meng}},\
  }\href {\doibase 10.1126/sciadv.1400260} {\bibfield  {journal} {\bibinfo
  {journal} {Sci. Adv.}\ }\textbf {\bibinfo {volume} {1}},\ \bibinfo {pages}
  {e1400260} (\bibinfo {year} {2015})}\BibitemShut {NoStop}%
\bibitem [{\citenamefont {Ohta}\ \emph {et~al.}(2012)\citenamefont {Ohta},
  \citenamefont {Cohen}, \citenamefont {Hirose}, \citenamefont {Haule},
  \citenamefont {Shimizu},\ and\ \citenamefont {Ohishi}}]{Ohta2012}%
  \BibitemOpen
  \bibfield  {author} {\bibinfo {author} {\bibfnamefont {K.}~\bibnamefont
  {Ohta}}, \bibinfo {author} {\bibfnamefont {R.~E.}\ \bibnamefont {Cohen}},
  \bibinfo {author} {\bibfnamefont {K.}~\bibnamefont {Hirose}}, \bibinfo
  {author} {\bibfnamefont {K.}~\bibnamefont {Haule}}, \bibinfo {author}
  {\bibfnamefont {K.}~\bibnamefont {Shimizu}}, \ and\ \bibinfo {author}
  {\bibfnamefont {Y.}~\bibnamefont {Ohishi}},\ }\href {\doibase
  10.1103/PhysRevLett.108.026403} {\bibfield  {journal} {\bibinfo  {journal}
  {Phys. Rev. Lett.}\ }\textbf {\bibinfo {volume} {108}},\ \bibinfo {pages}
  {026403} (\bibinfo {year} {2012})}\BibitemShut {NoStop}%
\bibitem [{\citenamefont {Leonov}(2015)}]{Leonov2015}%
  \BibitemOpen
  \bibfield  {author} {\bibinfo {author} {\bibfnamefont {I.}~\bibnamefont
  {Leonov}},\ }\href {\doibase 10.1103/PhysRevB.92.085142} {\bibfield
  {journal} {\bibinfo  {journal} {Phys. Rev. B}\ }\textbf {\bibinfo {volume}
  {92}},\ \bibinfo {pages} {085142} (\bibinfo {year} {2015})}\BibitemShut
  {NoStop}%
\bibitem [{\citenamefont {Badro}\ \emph {et~al.}(1999)\citenamefont {Badro},
  \citenamefont {Struzhkin}, \citenamefont {Shu}, \citenamefont {Hemley},
  \citenamefont {Mao}, \citenamefont {Kao}, \citenamefont {Rueff},\ and\
  \citenamefont {Shen}}]{Badro1999}%
  \BibitemOpen
  \bibfield  {author} {\bibinfo {author} {\bibfnamefont {J.}~\bibnamefont
  {Badro}}, \bibinfo {author} {\bibfnamefont {V.~V.}\ \bibnamefont
  {Struzhkin}}, \bibinfo {author} {\bibfnamefont {J.}~\bibnamefont {Shu}},
  \bibinfo {author} {\bibfnamefont {R.~J.}\ \bibnamefont {Hemley}}, \bibinfo
  {author} {\bibfnamefont {H.~K.}\ \bibnamefont {Mao}}, \bibinfo {author}
  {\bibfnamefont {C.~C.}\ \bibnamefont {Kao}}, \bibinfo {author} {\bibfnamefont
  {J.-P.}\ \bibnamefont {Rueff}}, \ and\ \bibinfo {author} {\bibfnamefont
  {G.}~\bibnamefont {Shen}},\ }\href {\doibase 10.1103/PhysRevLett.83.4101}
  {\bibfield  {journal} {\bibinfo  {journal} {Phys. Rev. Lett.}\ }\textbf
  {\bibinfo {volume} {83}},\ \bibinfo {pages} {4101} (\bibinfo {year}
  {1999})}\BibitemShut {NoStop}%
\bibitem [{\citenamefont {Kobayashi}\ \emph {et~al.}(2001)\citenamefont
  {Kobayashi}, \citenamefont {Takeshita}, \citenamefont {M{\^{o}}ri},
  \citenamefont {Takahashi},\ and\ \citenamefont {Kamimura}}]{Kobayashi2001}%
  \BibitemOpen
  \bibfield  {author} {\bibinfo {author} {\bibfnamefont {H.}~\bibnamefont
  {Kobayashi}}, \bibinfo {author} {\bibfnamefont {N.}~\bibnamefont
  {Takeshita}}, \bibinfo {author} {\bibfnamefont {N.}~\bibnamefont
  {M{\^{o}}ri}}, \bibinfo {author} {\bibfnamefont {H.}~\bibnamefont
  {Takahashi}}, \ and\ \bibinfo {author} {\bibfnamefont {T.}~\bibnamefont
  {Kamimura}},\ }\href {\doibase 10.1103/PhysRevB.63.115203} {\bibfield
  {journal} {\bibinfo  {journal} {Phys. Rev. B}\ }\textbf {\bibinfo {volume}
  {63}},\ \bibinfo {pages} {115203} (\bibinfo {year} {2001})}\BibitemShut
  {NoStop}%
\bibitem [{\citenamefont {Rueff}\ \emph {et~al.}(1999)\citenamefont {Rueff},
  \citenamefont {Kao}, \citenamefont {Struzhkin}, \citenamefont {Badro},
  \citenamefont {Shu}, \citenamefont {Hemley},\ and\ \citenamefont
  {Mao}}]{Rueff1999}%
  \BibitemOpen
  \bibfield  {author} {\bibinfo {author} {\bibfnamefont {J.~P.}\ \bibnamefont
  {Rueff}}, \bibinfo {author} {\bibfnamefont {C.~C.}\ \bibnamefont {Kao}},
  \bibinfo {author} {\bibfnamefont {V.~V.}\ \bibnamefont {Struzhkin}}, \bibinfo
  {author} {\bibfnamefont {J.}~\bibnamefont {Badro}}, \bibinfo {author}
  {\bibfnamefont {J.}~\bibnamefont {Shu}}, \bibinfo {author} {\bibfnamefont
  {R.~J.}\ \bibnamefont {Hemley}}, \ and\ \bibinfo {author} {\bibfnamefont
  {H.~K.}\ \bibnamefont {Mao}},\ }\href {{\textless}Go to
  ISI{\textgreater}://000083133200063} {\bibfield  {journal} {\bibinfo
  {journal} {Phys. Rev. Lett.}\ }\textbf {\bibinfo {volume} {83}},\ \bibinfo
  {pages} {3343} (\bibinfo {year} {1999})}\BibitemShut {NoStop}%
\bibitem [{\citenamefont {Fujimori}\ \emph {et~al.}(1996)\citenamefont
  {Fujimori}, \citenamefont {Mamiya}, \citenamefont {Mizokawa}, \citenamefont
  {Miyadai}, \citenamefont {Sekiguchi}, \citenamefont {Takahashi},
  \citenamefont {Mori},\ and\ \citenamefont {Suga}}]{Fujimori1996}%
  \BibitemOpen
  \bibfield  {author} {\bibinfo {author} {\bibfnamefont {A.}~\bibnamefont
  {Fujimori}}, \bibinfo {author} {\bibfnamefont {K.}~\bibnamefont {Mamiya}},
  \bibinfo {author} {\bibfnamefont {T.}~\bibnamefont {Mizokawa}}, \bibinfo
  {author} {\bibfnamefont {T.}~\bibnamefont {Miyadai}}, \bibinfo {author}
  {\bibfnamefont {T.}~\bibnamefont {Sekiguchi}}, \bibinfo {author}
  {\bibfnamefont {H.}~\bibnamefont {Takahashi}}, \bibinfo {author}
  {\bibfnamefont {N.}~\bibnamefont {Mori}}, \ and\ \bibinfo {author}
  {\bibfnamefont {S.}~\bibnamefont {Suga}},\ }\href
  {http://prb.aps.org/abstract/PRB/v54/i23/p16329{\_}1$\backslash$npapers2://publication/uuid/06022E99-2D72-4712-97F0-BB79D546921D}
  {\bibfield  {journal} {\bibinfo  {journal} {Phys. Rev. B}\ }\textbf {\bibinfo
  {volume} {54}},\ \bibinfo {pages} {16329} (\bibinfo {year}
  {1996})}\BibitemShut {NoStop}%
\bibitem [{\citenamefont {Chattopadhyay}\ and\ \citenamefont {von
  Schnering}(1985)}]{Chattopadhyay1985}%
  \BibitemOpen
  \bibfield  {author} {\bibinfo {author} {\bibfnamefont {T.}~\bibnamefont
  {Chattopadhyay}}\ and\ \bibinfo {author} {\bibfnamefont {H.~G.}\ \bibnamefont
  {von Schnering}},\ }\href {\doibase 10.1016/0022-3697(85)90204-5} {\bibfield
  {journal} {\bibinfo  {journal} {J. Phys. Chem. Solids}\ }\textbf {\bibinfo
  {volume} {46}},\ \bibinfo {pages} {113} (\bibinfo {year} {1985})}\BibitemShut
  {NoStop}%
\bibitem [{\citenamefont {Miyahara}\ and\ \citenamefont
  {Teranishi}(1968)}]{Miyahara1968}%
  \BibitemOpen
  \bibfield  {author} {\bibinfo {author} {\bibfnamefont {S.}~\bibnamefont
  {Miyahara}}\ and\ \bibinfo {author} {\bibfnamefont {T.}~\bibnamefont
  {Teranishi}},\ }\href {\doibase 10.1063/1.1656326} {\bibfield  {journal}
  {\bibinfo  {journal} {J. Appl. Phys.}\ }\textbf {\bibinfo {volume} {39}},\
  \bibinfo {pages} {896} (\bibinfo {year} {1968})}\BibitemShut {NoStop}%
\bibitem [{\citenamefont {Kune{\v{s}}}\ \emph {et~al.}(2010)\citenamefont
  {Kune{\v{s}}}, \citenamefont {Baldassarre}, \citenamefont {Sch{\"{a}}chner},
  \citenamefont {Rabia}, \citenamefont {Kuntscher}, \citenamefont {Korotin},
  \citenamefont {Anisimov}, \citenamefont {McLeod}, \citenamefont {Kurmaev},\
  and\ \citenamefont {Moewes}}]{Kunes2010}%
  \BibitemOpen
  \bibfield  {author} {\bibinfo {author} {\bibfnamefont {J.}~\bibnamefont
  {Kune{\v{s}}}}, \bibinfo {author} {\bibfnamefont {L.}~\bibnamefont
  {Baldassarre}}, \bibinfo {author} {\bibfnamefont {B.}~\bibnamefont
  {Sch{\"{a}}chner}}, \bibinfo {author} {\bibfnamefont {K.}~\bibnamefont
  {Rabia}}, \bibinfo {author} {\bibfnamefont {C.~A.}\ \bibnamefont
  {Kuntscher}}, \bibinfo {author} {\bibfnamefont {D.~M.}\ \bibnamefont
  {Korotin}}, \bibinfo {author} {\bibfnamefont {V.~I.}\ \bibnamefont
  {Anisimov}}, \bibinfo {author} {\bibfnamefont {J.~A.}\ \bibnamefont
  {McLeod}}, \bibinfo {author} {\bibfnamefont {E.~Z.}\ \bibnamefont {Kurmaev}},
  \ and\ \bibinfo {author} {\bibfnamefont {A.}~\bibnamefont {Moewes}},\ }\href
  {\doibase 10.1103/PhysRevB.81.035122} {\bibfield  {journal} {\bibinfo
  {journal} {Phys. Rev. B}\ }\textbf {\bibinfo {volume} {81}},\ \bibinfo
  {pages} {035122} (\bibinfo {year} {2010})}\BibitemShut {NoStop}%
\bibitem [{\citenamefont {Moon}\ \emph {et~al.}(2015)\citenamefont {Moon},
  \citenamefont {Kang}, \citenamefont {Jang},\ and\ \citenamefont
  {Shim}}]{Moon2015}%
  \BibitemOpen
  \bibfield  {author} {\bibinfo {author} {\bibfnamefont {C.-Y.}\ \bibnamefont
  {Moon}}, \bibinfo {author} {\bibfnamefont {H.}~\bibnamefont {Kang}}, \bibinfo
  {author} {\bibfnamefont {B.~G.}\ \bibnamefont {Jang}}, \ and\ \bibinfo
  {author} {\bibfnamefont {J.~H.}\ \bibnamefont {Shim}},\ }\href {\doibase
  10.1103/PhysRevB.92.235130} {\bibfield  {journal} {\bibinfo  {journal} {Phys.
  Rev. B}\ }\textbf {\bibinfo {volume} {92}},\ \bibinfo {pages} {235130}
  (\bibinfo {year} {2015})}\BibitemShut {NoStop}%
\bibitem [{\citenamefont {Kotliar}\ and\ \citenamefont
  {Vollhardt}(2004)}]{Kotliar2004}%
  \BibitemOpen
  \bibfield  {author} {\bibinfo {author} {\bibfnamefont {G.}~\bibnamefont
  {Kotliar}}\ and\ \bibinfo {author} {\bibfnamefont {D.}~\bibnamefont
  {Vollhardt}},\ }\href {\doibase 10.1063/1.1712502} {\bibfield  {journal}
  {\bibinfo  {journal} {Phys.Today}\ }\textbf {\bibinfo {volume} {57}},\
  \bibinfo {pages} {53} (\bibinfo {year} {2004})}\BibitemShut {NoStop}%
\bibitem [{\citenamefont {Blaha}\ \emph {et~al.}(2001)\citenamefont {Blaha},
  \citenamefont {Schwarz},\ and\ \citenamefont {Madsen}}]{Blaha2001}%
  \BibitemOpen
  \bibfield  {author} {\bibinfo {author} {\bibfnamefont {P.}~\bibnamefont
  {Blaha}}, \bibinfo {author} {\bibfnamefont {K.}~\bibnamefont {Schwarz}}, \
  and\ \bibinfo {author} {\bibfnamefont {G.}~\bibnamefont {Madsen}},\ }\href
  {\doibase citeulike-article-id:6205108} {\emph {\bibinfo {title} {Isbn
  3-9501031-1-2}}}\ (\bibinfo {year} {2001})\ p.\ \bibinfo {pages}
  {2001}\BibitemShut {NoStop}%
\bibitem [{\citenamefont {Perdew}\ \emph {et~al.}(1997)\citenamefont {Perdew},
  \citenamefont {Burke},\ and\ \citenamefont {Ernzerhof}}]{Perdew1997}%
  \BibitemOpen
  \bibfield  {author} {\bibinfo {author} {\bibfnamefont {J.~P.}\ \bibnamefont
  {Perdew}}, \bibinfo {author} {\bibfnamefont {K.}~\bibnamefont {Burke}}, \
  and\ \bibinfo {author} {\bibfnamefont {M.}~\bibnamefont {Ernzerhof}},\ }\href
  {\doibase 10.1103/PhysRevLett.78.1396} {\bibfield  {journal} {\bibinfo
  {journal} {Phys. Rev. Lett.}\ }\textbf {\bibinfo {volume} {78}},\ \bibinfo
  {pages} {1396} (\bibinfo {year} {1997})}\BibitemShut {NoStop}%
\bibitem [{\citenamefont {Haule}\ \emph {et~al.}(2010)\citenamefont {Haule},
  \citenamefont {Yee},\ and\ \citenamefont {Kim}}]{Haule2010}%
  \BibitemOpen
  \bibfield  {author} {\bibinfo {author} {\bibfnamefont {K.}~\bibnamefont
  {Haule}}, \bibinfo {author} {\bibfnamefont {C.~H.}\ \bibnamefont {Yee}}, \
  and\ \bibinfo {author} {\bibfnamefont {K.}~\bibnamefont {Kim}},\ }\href
  {http://journals.aps.org/prb/abstract/10.1103/PhysRevB.81.195107} {\bibfield
  {journal} {\bibinfo  {journal} {Phys. Rev. B}\ }\textbf {\bibinfo {volume}
  {81}},\ \bibinfo {pages} {195107} (\bibinfo {year} {2010})}\BibitemShut
  {NoStop}%
\bibitem [{\citenamefont {Kresse}\ and\ \citenamefont
  {Furthm{\"{u}}ller}(1996)}]{Kresse1996}%
  \BibitemOpen
  \bibfield  {author} {\bibinfo {author} {\bibfnamefont {G.}~\bibnamefont
  {Kresse}}\ and\ \bibinfo {author} {\bibfnamefont {J.}~\bibnamefont
  {Furthm{\"{u}}ller}},\ }\href {\doibase 10.1103/PhysRevB.54.11169} {\bibfield
   {journal} {\bibinfo  {journal} {Phys. Rev. B}\ }\textbf {\bibinfo {volume}
  {54}},\ \bibinfo {pages} {11169} (\bibinfo {year} {1996})}\BibitemShut
  {NoStop}%
\bibitem [{\citenamefont {Imada}\ \emph {et~al.}(1998)\citenamefont {Imada},
  \citenamefont {Fujimori},\ and\ \citenamefont {Tokura}}]{Imada1998}%
  \BibitemOpen
  \bibfield  {author} {\bibinfo {author} {\bibfnamefont {M.}~\bibnamefont
  {Imada}}, \bibinfo {author} {\bibfnamefont {A.}~\bibnamefont {Fujimori}}, \
  and\ \bibinfo {author} {\bibfnamefont {Y.}~\bibnamefont {Tokura}},\ }\href
  {\doibase 10.1103/RevModPhys.70.1039} {\bibfield  {journal} {\bibinfo
  {journal} {Rev. Mod. Phys.}\ }\textbf {\bibinfo {volume} {70}},\ \bibinfo
  {pages} {1039} (\bibinfo {year} {1998})}\BibitemShut {NoStop}%
\end{thebibliography}%

\end{document}